\documentclass{article}
\usepackage{spconf,amsmath,graphicx,hyperref}
\usepackage{cite} 
\usepackage{adjustbox}
\usepackage{booktabs}
\usepackage{subcaption}
\usepackage{array}
\usepackage{comment}
\usepackage{color}
\title{Entropy-Guided GRVQ for Ultra-Low Bitrate Neural Speech Codec}

\name{
Yanzhou Ren$^{*}$,
Noboru Harada$^{\dagger}$,
Daiki Takeuchi$^{\dagger}$,
Siyu Chen$^{*}$,
Wei Liu$^{*}$,
Xiao Zhang$^{*}$,
Liyuan Zhang$^{*}$, 
Takehiro Moriya$^{\dagger}$,
and Shoji Makino$^{*}$
}

\address{
$^{*}$ Waseda University, Japan \\
$^{\dagger}$ NTT, Inc., Japan
}

\begin{document}
\ninept
\maketitle

\begin{abstract}
Neural audio codec (NAC) is essential for reconstructing high-quality speech signals and generating discrete representations for downstream speech language models. However, ensuring accurate semantic modeling while maintaining high-fidelity reconstruction under ultra-low bitrate constraints remains challenging. We propose an entropy-guided group residual vector quantization (EG-GRVQ) for an ultra-low bitrate neural speech codec, which retains a semantic branch for linguistic information and incorporates an entropy-guided grouping strategy in the acoustic branch. Assuming that channel activations follow approximately Gaussian statistics, the variance of each channel can serve as a principled proxy for its information content. Based on this assumption, we partition the encoder output such that each group carries an equal share of the total information.
This balanced allocation improves codebook efficiency and reduces redundancy. Trained on LibriTTS and VCTK, our model shows improvements in perceptual quality and intelligibility metrics under ultra-low bitrate conditions, with a focus on codec-level fidelity for communication-oriented scenarios.
\end{abstract}
\vspace{-3mm}
\section{Introduction}
\label{sec:intro}

Neural audio codec (NAC) has gained increasing importance in modern speech processing by producing discrete representations that serve different purposes in speech applications~\cite{van2017neural,yang2021source,yang2024towards,chen2025neural,wu2023audiodec,borsos2023soundstorm}.
Prior studies categorize discrete speech representations into two main types: semantic tokens and acoustic tokens~\cite{borsos2023audiolm}. Semantic tokens capture high-level linguistic information and are widely adopted in downstream speech-language modeling tasks~\cite{brychcin2014semantic}. They are commonly extracted from self-supervised pretrained models such as HuBERT and WavLM~\cite{baevski2020wav2vec,chen2022wavlm}. They can subsequently be incorporated into neural codecs through semantic-aware quantization~\cite{liu2024semanticodec,zhang2024speechtokenizer,defossez2024moshi}, enabling accurate content alignment for applications such as speech-to-text, speech translation, and zero-shot text-to-speech~\cite{hsu2021hubert,zhu2025vec,wang2024maskgct,zhang2023speechgpt}.
Acoustic tokens, typically obtained from neural audio codecs, are designed to preserve fine-grained signal details for waveform reconstruction. Neural codecs such as SoundStream~\cite{zeghidour2021soundstream}, EnCodec~\cite{defossez2022high}, and Stable Codec~\cite{parker2024scaling} rely on residual vector quantization (RVQ) or finite scale quantization (FSQ), while HIFI-Codec~\cite{yang2023hifi} further introduced grouped residual vector quantization (GRVQ) to improve perceptual quality by distributing information across quantization groups. 
It is worth noting that the vector quantization of coefficients for ultra-low bit-rate codecs was an essential research topic over 35 years ago~\cite{shiraki1988lpc}, and the similar topic is now being revisited in the context of neural speech and audio coding.

Although semantic and acoustic tokens provide complementary benefits, ensuring both under ultra-low bitrate constraints remains an open challenge. Recent works have attempted to integrate both aspects within a single codec. For example, Mimi~\cite{defossez2024moshi} introduces a parallel quantization design with dedicated semantic and acoustic information branches. It utilizes an RVQ to code acoustic information. It also allocates bitrate capacity to semantic tokens, which inevitably reduces the resources for acoustic details~\cite{ji2024wavtokenizer}, thus constraining simultaneous improvement of intelligibility and reconstruction quality, especially for ultra-low bitrate conditions.

In this paper, we adopt Mimi~\cite{defossez2024moshi} as our baseline, for its causal, low-latency architecture suitable for real-time communication scenarios, with particular emphasis on strengthening the acoustic branch to achieve higher reconstruction fidelity. We propose an entropy-guided group residual vector quantization (EG-GRVQ) strategy, which extends the GRVQ framework originally introduced in HiFi-Codec~\cite{yang2023hifi}.
 The EG-GRVQ is motivated by the information-theoretic link between variance and entropy, where channel variance serves as a proxy for information content. Encoder outputs are partitioned to equalize cumulative variance across groups, leading to a more balanced allocation of information and improved codebook efficiency. While explicitly retaining Mimi’s semantic branch, our design ensures that the acoustic branch achieves higher fidelity under low-bitrate conditions. Experiments on LibriTTS~\cite{zen2019libritts} and VCTK~\cite{yamagishi2019cstr} confirm that our approach yields improvements in both perceptual quality and intelligibility.

\section{PROPOSED METHOD}
\begin{figure*}[t]
  \centering
  \includegraphics[width= 0.88 \textwidth]{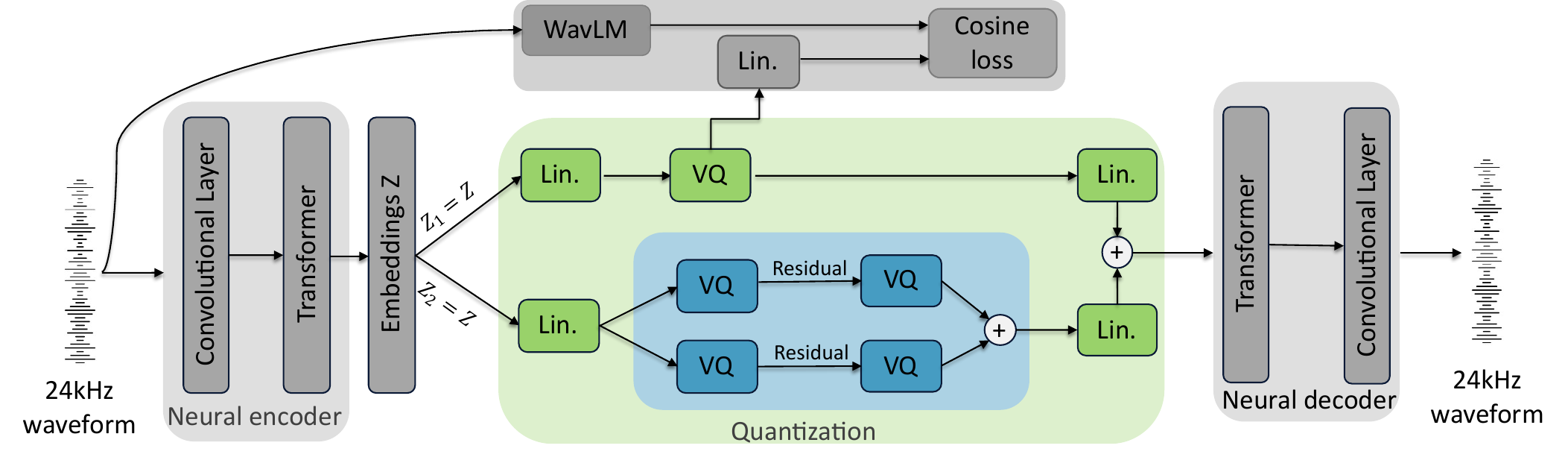}
  \caption{Structure of the proposed model.}
  % \vspace*{-3pt}
 % {\hfill\footnotesize Note how the caption is centered in the column.\hfill}
  \label{fig:magnetization}
\end{figure*}

\subsection{Encoder and Decoder}

The encoder and decoder of our proposal follow the overall architecture of Mimi~\cite{defossez2024moshi}. As shown in Fig.~\ref{fig:magnetization}, the input is 24 kHz waveforms, which are first passed through four residual convolutional blocks and finally a 1D convolution in the encoder. These blocks progressively reduce the temporal resolution and increase the channel dimension, ultimately transforming the waveform into 512-dimensional latent representations at a rate of 12.5 frames per second. 
Following the convolutional blocks, Transformer blocks are applied in the bottleneck to capture long-range dependencies and enhance the compactness of latent representations.
The decoder adopts a symmetric architecture to the encoder, operating in reverse to reconstruct the waveform. It takes the quantized latent features as input and reconstructs the waveform through the transposed convolutional blocks that progressively upsample the temporal resolution and reduce the channel dimensions.

\subsection{Quantization}

As shown in Fig.~\ref{fig:magnetization}, after encoding, the latent features are duplicated into two branches. The first branch is processed by a semantic quantizer, designed to capture high-level linguistic features that are critical for speech intelligibility. This semantic layer employs a single codebook to minimize bitrate consumption while still preserving essential semantic information.  

\begin{figure}[t]
  \centering
  \includegraphics[width=1.0\linewidth]{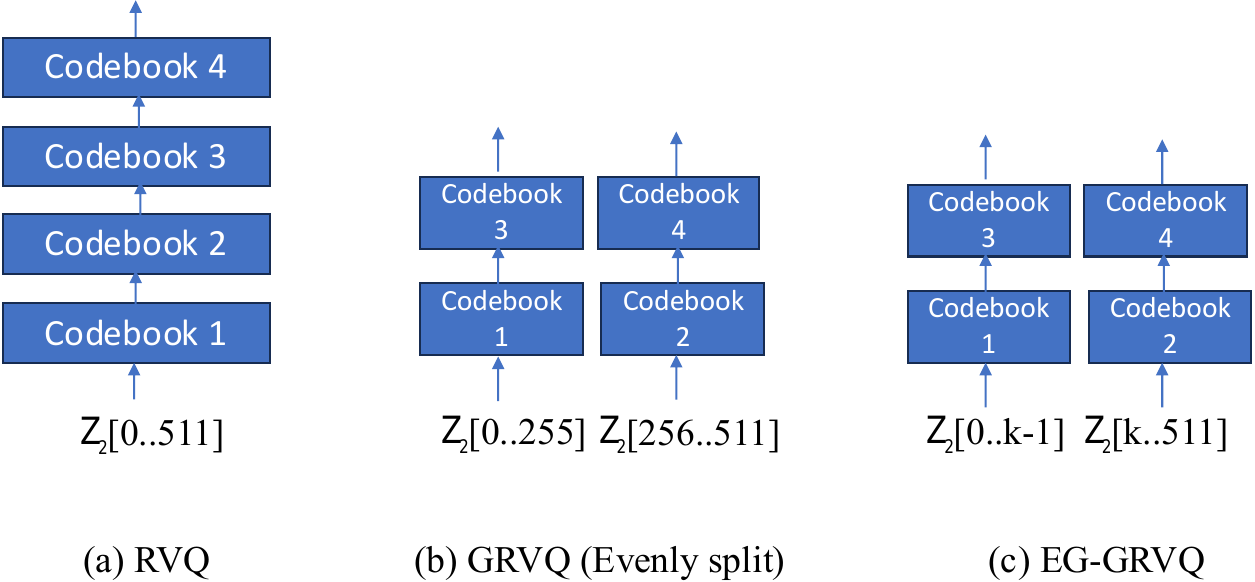}
  %\vspace{-4mm}
  \caption{Quantizer structure configuration for (a) RVQ, (b) GRVQ, and (c) EG-GRVQ (Proposal).}
  \label{fig:structure}
\end{figure}

The second branch is processed by an entropy-guided grouped residual vector quantizer. Fig.~\ref{fig:structure} shows the three different quantizer structure configurations for acoustic information: (a) RVQ, (b) GRVQ and (c) EG-GRVQ (Proposal). While RVQ sequentially utilizes a series of residual codebooks to quantize the target acoustic vector $Z_2$, the conventional GRVQ splits the channels of the acoustic target vector $Z_2$ into two coding groups.

Unlike conventional GRVQ approaches that divide channels evenly, our method, EG-GRVQ, exploits the global variance distribution of encoder outputs to guide grouping.  Let $z_{k,t}$ denote the encoder activation at channel $k \in \{1,\dots,C\}$ and time frame $t \in \{1,\dots,T\}$. The variance of channel $k$ over the training set is defined as
\begin{equation}
\sigma_k^2 = \frac{1}{T}\sum_{t=1}^{T} \big(z_{k,t} - \mu_k\big)^2,
\end{equation}
where $\mu_k$ is the average output of channel $k$.  

From an information-theoretic perspective, we assume that each encoder channel, after subtracting its mean constant bias, approximately follows a Gaussian distribution~\cite{jacot2018neural}. 
Accordingly, the differential entropy is given by
\begin{equation}
H(X) = \tfrac{1}{2}\ln \big(2\pi e \sigma^2\big),
\end{equation}
indicating that entropy is a monotonic function of variance. Hence, channel variance can serve as a proxy for the amount of information carried by each channel.  

We then determine the smallest index $k^\ast$ such that
\begin{equation}
\sum_{i=1}^{k^\ast} \sigma_i^2 \;\;\geq\;\; \tfrac{1}{2} \sum_{j=1}^{C} \sigma_j^2,
\end{equation}
ensuring that the first group accounts for approximately half of the total variance. 
In our implementation with $C=512$, the split point is $k=237$, producing Group~1 (Codebooks 1 and 3 as shown in Fig.~\ref{fig:structure}-(c)) with 237 channels and Group~2 (Codebooks 2 and 4 as shown in Fig.~\ref{fig:structure}-(c)) with 275 channels. The even split at
$k = 256$ allocates 55.30 \% of the variance to the first half and 44.70 \% to the second half. This strategy mitigates the dominance of high-variance channels and the under-utilization of low-variance channels, thereby enhancing quantization efficiency and reconstruction fidelity. 

Note that when some signaling bits are allocated, the split position $k$ can be adaptive in each frame. However, considering the trade-off for spending additional bits and coding gain improvement, no adaptation has been applied in this proposal. The resulting grouping remains fixed as a hyperparameter for all data during both training and inference.

Finally, the quantized features from semantic branch and acoustic branch are summed and passed to the decoder for waveform reconstruction.  

%%%%%%% Table 1 %%%%%%%%
\begin{table*}[th!]
\centering
\caption{Objective evaluation results at 0.6875 kbps.}
\label{tab:main_results}
% \vspace{-6pt}
\begin{tabular}{lcccccc}
\toprule
Method & VQ scheme (Structure) & SDR↑ & PESQ↑ & STOI↑ & ViSQOL↑ \\
\midrule
Mimi (official) & RVQ (1x4)  & \textbf{3.451}  & 1.872 & 0.876 & 2.010 \\
Mimi (retrain) & RVQ (1x4) & -6.969 & 1.779 & 0.886 & \textbf{2.546} \\
Mimi (GRVQ) & GRVQ (2x2) & -7.294 & 1.852 & 0.889 & 2.464 \\
%Comparison 1 & Split VQ (4x1) & -7.420 & 1.736 & 0.875 & 2.026 \\
%Comparison 2 & GRVQ (2x2) & -7.294 & 1.852 & 0.889 & 2.464 \\
Proposal & EG-GRVQ (2x2) & -7.309 & \textbf{1.881} & \textbf{0.890} & 2.496 \\
\bottomrule
\end{tabular}
\end{table*}

%%%%%%% Table 2 %%%%%%%%
\begin{table*}[th!]
\centering
\caption{Normalized Mean Squared Error (NMSE) across acoustic quantizers.}
\label{tab:nmse}
% \vspace{-6pt}
\begin{adjustbox}{max width=\textwidth}
\begin{tabular}{lccccc}
\toprule
Method & Codebook1 & Codebook2 & Codebook3 & Codebook4 & Total \\
\midrule
RVQ            & 0.928 & 0.908 & 0.894 & 0.884 & 0.884 \\
GRVQ           & 0.890 & 0.895 & 0.842 & 0.862 & 0.852 \\
Proposal (EG-GRVQ) & \textbf{0.865} & \textbf{0.855} & \textbf{0.831} & \textbf{0.813} & \textbf{0.819} \\
\bottomrule
\end{tabular}
\end{adjustbox}
\end{table*}

\section{Experiments}
\label{sec:Experiments}

\subsection{Baselines}
In our experiments, we consider four systems as baselines for comparison. 
We denote the original Mimi codec without retraining as Mimi (official), 
the retrained version with residual vector quantization as Mimi (retrain), 
and our re-implementation of grouped residual vector quantization within the Mimi 
framework as Mimi (GRVQ). Our proposed method, based on entropy-guided 
grouping, is referred to as Proposal (EG-GRVQ).
\subsection{Datasets and Experimental Setup}
We trained our proposed method on the combined LibriTTS~\cite{zen2019libritts} and VCTK~\cite{yamagishi2019cstr}, including the train-clean-100, train-clean-360, and train-other-500 subsets of LibriTTS, as well as the full VCTK corpus. To ensure a fair and objective comparison, we retrained the Mimi baseline under the same data conditions.
For comparison, we also re-implemented the previously proposed GRVQ quantization~\cite{yang2023hifi} within the Mimi framework, denoted as Mimi (GRVQ).
All trainings and evaluations were conducted using the 5-codebook setting under the same conditions.
All models were trained on 8 NVIDIA A6000 GPUs (48 GB each), with a batch size of 12 per GPU. 
\subsection{Training Strategy}
To train the proposed model, we adopt a multiobjective strategy that integrates feature loss and adversarial loss, with a discriminator-based scheme to enhance perceptual quality. The generator is optimized by minimizing the following composite loss:
\begin{equation}
\mathcal{L}_{\text{gen}} = \lambda_{\text{adv}} \mathcal{L}_{\text{adv}} + \lambda_{\text{feat}} \mathcal{L}_{\text{FM}},
\end{equation}
where:
\begin{itemize}
    \item $\mathcal{L}_{\text{adv}}$ is the adversarial loss, computed as the mean squared error (MSE) between discriminator predictions and the target label $1$ (real):
    \begin{equation}
    \mathcal{L}_{\text{adv}} = \frac{1}{N} \sum_{i=1}^{N}(D(\hat{x}_i) - 1)^2.
    \end{equation}
    \item $\mathcal{L}_{\text{FM}}$ is the feature matching loss, defined as the $L_1$ distance between real and generated intermediate discriminator features:
    \begin{equation}
    \mathcal{L}_{\text{FM}} = \frac{1}{N}
    \sum_{l=1}^{N}
    \left\lVert D^{(l)}(x) - D^{(l)}(\hat{x}) \right\rVert_1,
    \end{equation}
    where $N$ is the total number of feature maps across all layers and scales.
\end{itemize}

We set $\lambda_{\text{adv}} = 1$ and $\lambda_{\text{feat}} = 15$. In addition, a vector quantization commitment loss $\mathcal{L}_{\text{commit}}$, weighted by $\lambda_{\text{commit}} = 1$, is included to stabilize codebook usage.

For semantic learning, we followed Speechtokenizer~\cite{zhang2024speechtokenizer} by using a distillation loss between projected semantic quantizer outputs and WavLM-derived embeddings. The latent features from the semantic quantizer are projected to a 1024-dimensional space and then compared with WavLM-derived semantic embeddings using a cosine similarity loss ${L}_{\text{sem}}$. The semantic distillation loss ${L}_{\text{sem}}$ is computed as:

\begin{equation}
\mathcal{L}_{\text{sem}} = 1 - \cos(z_s, z_{\text{WavLM}}).
\end{equation}

\subsection{Objective Evaluation}

For objective evaluation, we randomly selected 200 samples from the LibriTTS~\cite{zen2019libritts} test-clean subset, which was strictly excluded from training. To prevent information leakage, we ensured that there was no data overlap between the training and evaluation sets. We employ four objective metrics to assess the reconstruction quality: Perceptual Evaluation of Speech Quality (PESQ)~\cite{rix2001perceptual}, Short-Time Objective Intelligibility (STOI)~\cite{taal2010short}, Virtual Speech Quality Objective Listener (ViSQOL)~\cite{hines2015visqol}, and Signal-to-Distortion Ratio (SDR)~\cite{le2019sdr}.
Specifically, we adopt the wideband version of PESQ for perceptual quality, STOI for intelligibility under low-bitrate conditions, and ViSQOL for reference-based perceptual correlation. 
SDR, although widely used in separation and enhancement tasks, is sensitive to phase misalignment; therefore, we report it only as a supplementary metric.
During evaluation, all model outputs, originally at 24 kHz and 16-bit resolution, are downsampled to 16 kHz using a 7.5 kHz low-pass filter to meet the input requirements of PESQ and STOI.

As shown in Table~\ref{tab:main_results}, our proposed EG-GRVQ achieves the best performance in
PESQ and STOI, while maintaining competitive results in ViSQOL. 
Compared with Mimi (official), our method improves PESQ by
0.01, ViSQOL by 0.49, and STOI by 0.01, demonstrating substantial
gains in perceptual quality. Relative to Mimi (retrain), which
is trained on the same data, our method achieves an increase of 0.1
in PESQ and 0.01 in STOI while yielding comparable SDR 
(-7.31 vs. -6.97). Compared with Mimi (GRVQ), our method
further improves PESQ from 1.85 to 1.88 and ViSQOL from 2.46
to 2.50. These results validate the effectiveness of the proposed
entropy-guided design in enhancing perceptual quality under low
bitrate conditions.

In addition, we evaluate the normalized mean squared error (NMSE) between the quantized reconstruction and the oracle encoder output, which provides a scale-invariant measure of distortion. Table~\ref {tab:nmse} shows the NMSE results across acoustic quantizers. The proposed method consistently achieves lower NMSE, confirming that our grouping strategy yields more accurate reconstruction relative to latent variance.

\begin{figure}[t]
  \centering
  \includegraphics[width= 1\linewidth]{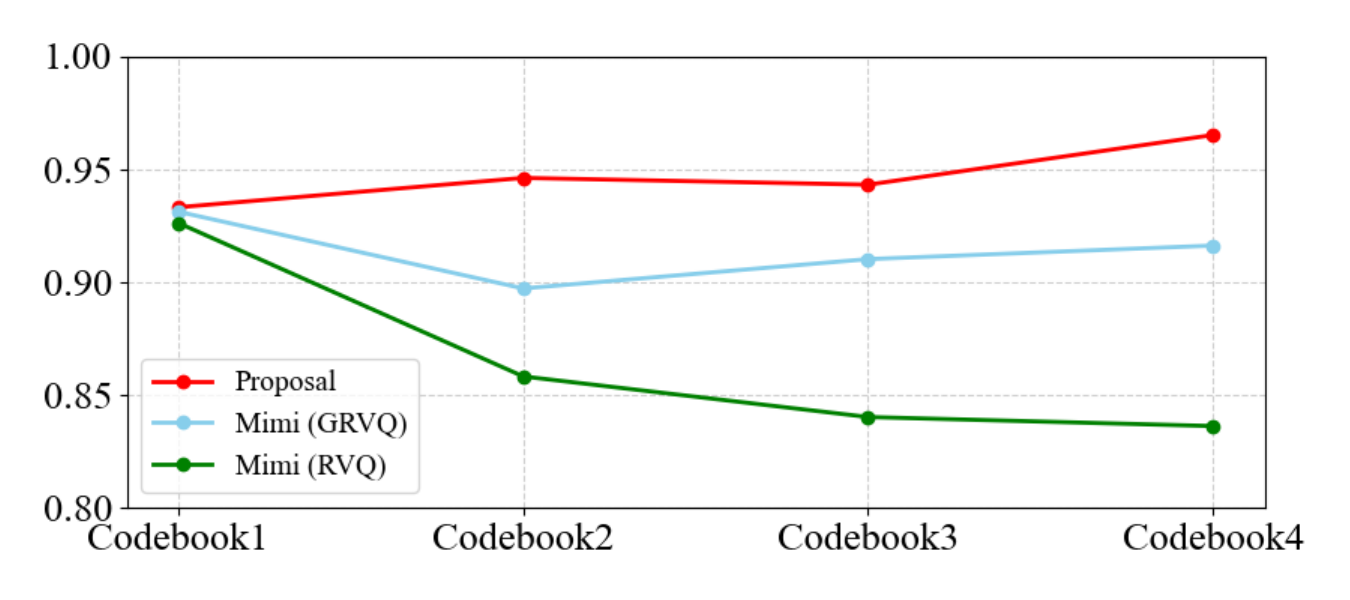}
  \caption{Codebook utilization rate in acoustic branch.}
  \label{fig:codebook}
\end{figure}

\subsection{Codebook Efficiency and Grouping Analyses}

We compare codebook utilization across the acoustic branch to evaluate the effectiveness of different quantization strategies. As shown in Fig.~\ref{fig:codebook}, the RVQ baseline exhibits a sharp decline in utilization in deeper layers, indicating inefficient codeword usage and potential room for improvement. The Mimi (GRVQ) baseline alleviates this problem to some extent by balancing the allocation across groups, but still suffers from layer-wise imbalance. In contrast, the proposed EG-GRVQ maintains consistently high and stable utilization across all layers, demonstrating more effective use of quantizer capacity. This indicates that our entropy-guided grouping not only preserves reconstruction accuracy but also maximizes representation efficiency under constrained bitrate.
% \textcolor{blue}{We observe that while all methods share similar codebook utilization at the first layer, the vanilla RVQ baseline suffers from a gradual decline in later layers. This degradation is likely due to the increasing difficulty of optimizing high-dimensional residual spaces. In contrast, our entropy-guided grouping mitigates this issue by dividing the latent space into simpler subspaces, which are easier to train. As a result, codebook usage remains consistently high across all layers. A slight increase in later stages is also observed, which we attribute to residual randomness becoming more visible when collapse is avoided.}

\begin{table}[tb]
\centering
\caption{Objective results of different grouping numbers at 0.6875 kbps.}
\label{tab:group_ablation}
% \vspace{-6pt}
\begin{tabular}{lcccc}
\toprule
Method & SDR↑ & PESQ↑ & STOI↑ & ViSQOL↑ \\
\midrule
1×4 (RVQ) & \textbf{-6.969} & 1.779 & 0.886 & \textbf{2.546} \\
4×1       & -7.420 & 1.736 & 0.875 & 2.026 \\
2×2 (GRVQ) & -7.294 & \textbf{1.852} & \textbf{0.889} & 2.464 \\
\bottomrule
\end{tabular}
\end{table}

To evaluate the impact of grouping numbers in the acoustic branch, we compared three configurations under a fixed total of 4 quantizers: $2 groups \times 2 quantizers, 4 groups \times 1 quantizer$, $1 group \times 4 quantizers$. As shown in Table~\ref{tab:group_ablation}, the $2 \times 2$
 configuration consistently outperforms the $4\times1$ variant across all the metrics. The 2-group structure enables deeper quantization per group, leading to better reconstruction quality, particularly in perceptual similarity. These results indicate that, under fixed bitrate, grouping fewer but deeper quantizers is more effective than shallowly quantizing more groups. Compared to the $1\times4$ configuration, the $2\times2$ variant achieves significantly higher PESQ and STOI, with equal ViSQOL but slightly lower SDR. This suggests that simply stacking quantizers in a single group may cause inefficiencies in information allocation, leading to degraded perceptual quality.
These results indicate that, under a fixed bitrate, grouping fewer but deeper quantizers is more effective than both shallow multi-group quantization and single-group quantization without structural separation.

\subsection{Subjective Evaluation}
In addition to objective metrics, we conducted the Multiple Stimuli with Hidden Reference and Anchor (MUSHRA) test to evaluate perceptual quality from a human listener's perspective. MUSHRA is widely adopted in audio coding and speech enhancement research. It provides a reliable subjective assessment by asking participants to rate the quality of multiple signal variants relative to a hidden reference and an explicitly degraded anchor. 

We randomly selected eight speech samples, four male and four female, from the LibriTTS test-clean subset, with durations ranging from 1.4 to 13 seconds. These speech samples were recorded at 24 kHz with 16-bit resolution for evaluation. 
The tested systems included an oracle clean reference, a low-pass anchor created by applying a 3.4 kHz low-pass filter to the reference, Mimi (GRVQ), Proposal (EG-GRVQ), and Mimi (retrain) and Mimi (official). 
The listening test was conducted with eight participants (five male and three female), all of whom had prior experience in perceptual audio evaluation.

\begin{figure}[t]
  \centering
  \includegraphics[width=0.88\linewidth]{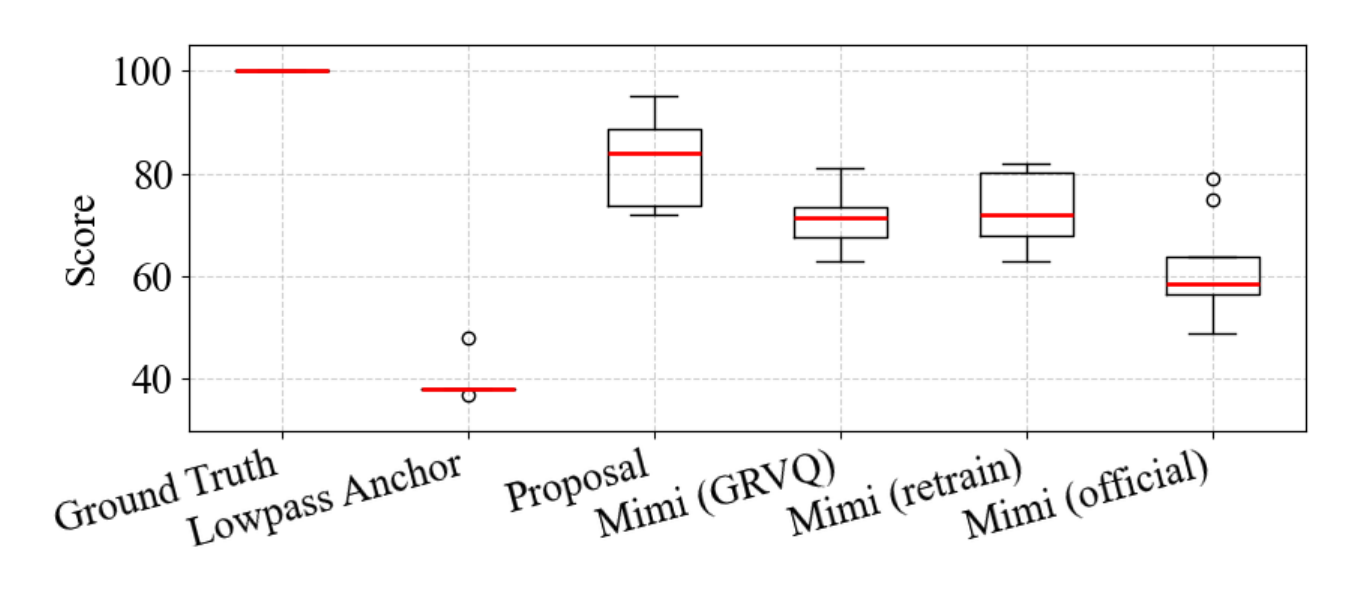}
  \caption{MUSHRA score distributions of different methods.}
  \label{fig:boxplot_method}
\end{figure}

\begin{figure}[t]
  \centering
  \includegraphics[width=0.88\linewidth]{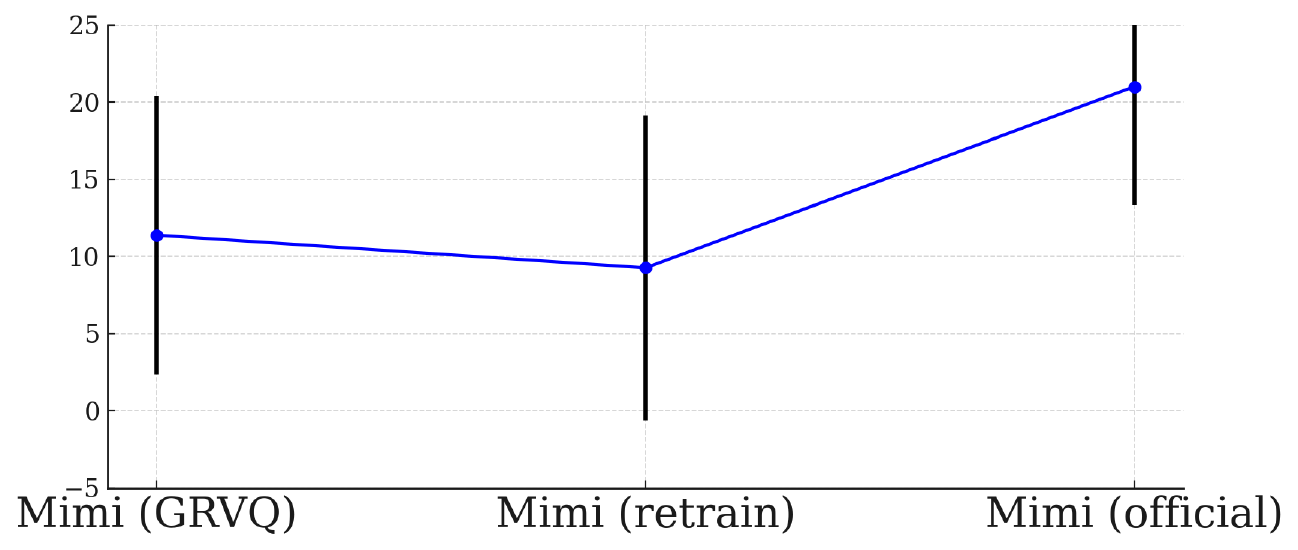}
  \caption{Mean MUSHRA score differences between Proposal (EG-GRVQ) and baselines with 95\% confidence intervals}
  \label{fig:CI}
\end{figure}

As shown in Fig.~\ref{fig:boxplot_method}, the proposed method significantly improves the MUSHRA score by 21 points over the official Mimi, and by 11 points over Mimi with GRVQ, indicating strong subjective preference. Fig.~\ref{fig:CI} further shows the mean differential MUSHRA scores with 95\% confidence intervals. The results indicate that the proposed method achieves statistically significant improvements over the official baseline and Mimi (GRVQ), as the CI ranges do not overlap with zero. These findings highlight that our proposed method offers clear advantages in perceptual preference under low-bitrate conditions.

\section{Conclusions}
\label{sec:Conclusions}
% \begin{table}[t]
% \centering
% \caption{Objective results of different grouping strategies at 0.6875 kbps.}
% \label{tab:strategy_ablation}
% \begin{tabular}{lcccc}
% \toprule
% \textbf{Split Method} & \textbf{SDR}↑ & \textbf{PESQ}↑ & \textbf{STOI}↑ & \textbf{ViSQOL}↑ \\
% \midrule
% No split & -7.280 & 1.795 & 0.880 & 2.225 \\
% GRVQ (Evenly split) & -7.294 & 1.852 & 0.889 & 2.464 \\
% \textbf{Proposal (EG-GRVQ)} & -7.309 & \textbf{1.881} & \textbf{0.890} & \textbf{2.496} \\
% \bottomrule
% \end{tabular}
% \end{table}

In this paper, we propose an entropy-guided group residual vector quantization framework, which demonstrates clear advantages for ultra-low bitrate neural speech coding. By leveraging channel variance as a proxy for entropy, our method ensures a more balanced allocation of information across quantization groups, thereby enhancing both reconstruction fidelity and codebook utilization rate. Objective evaluations show consistent improvements in PESQ and STOI while maintaining competitive ViSQOL and SDR scores, and NMSE analysis further confirms more accurate reconstruction relative to latent variance. Subjective MUSHRA tests also indicate a significant perceptual preference over baseline systems. Moreover, analysis of codebook usage reveals that the proposed design increased codebook utilization rate, leading to more efficient representation under constrained bitrate conditions. These results highlight the potential of EG-GRVQ for ultra-low bitrate neural speech codecs.

\newpage
% \footnotesize
\bibliographystyle{IEEEtran}
\bibliography{refs}

\end{document}